\documentclass[12pt,a4paper,fleqn]{article}
\usepackage{amsmath,amssymb,amsfonts,amsthm,amsopn,graphicx}
\usepackage{tikz} 
\usepackage{mathrsfs} 
\usetikzlibrary{arrows}
\usepackage{hyperref}

\usepackage[T1,T2A]{fontenc}
\allowdisplaybreaks[3]
\numberwithin{equation}{section} 
\newtheorem{theorem}{Theorem}[section]

\newtheorem{stat}[theorem]{Statement}

\newcommand{\ds}{\displaystyle}
\def\EXP{\textrm{{\large e}}}
\newcommand{\ii}{{\mathtt{i}}}

\newcommand{\ph}{\varphi}

\newcommand{\sk}[1]{\langle #1\rangle}

\newcommand{\M}{\mathsf{M}}
\renewcommand{\aa}{{\boldsymbol{a}}}
\newcommand{\cc}{{\boldsymbol{c}}}
\newcommand{\dd}{{\boldsymbol{d}}}
\newcommand{\ee}{{\boldsymbol{e}}}
\newcommand{\bb}{{\boldsymbol{b}}}

\newcommand{\bl}{{\mathtt{b}}}

\def\uop{\boldsymbol{u}}

\def\vop{\boldsymbol{v}}

\usepackage[body={15.5cm,22cm}]{geometry}
\begin{document}

\title{``Pentagonal Algebra'' and Four-Simplex Equation.}
\author{Sergey Sergeev
\thanks{    
    Faculty of Science and Technology, 
   University of Canberra, Bruce ACT 2617, Australia
}
\thanks{ 
Department of Fundamental and Theoretical Physics,
         Research School of Physics,
    Australian National University, Canberra, ACT 2601, Australia}
}

\date{}

\maketitle

\begin{abstract}
Some idea, which leads to a non-trivial solution of the quantum four-simplex equation, is exposed in this paper. We call this idea ``pentagonal algebra''. Few examples of the realisation of this idea are given here, and thus few examples of $R$-matrix for the quantum four-simplex equation are presented.
\end{abstract}


\tableofcontents


\section{Introduction}

The famous Yang-Baxter, or triangle, equation has a sequence of its multidimensional extensions: the tetrahedron equation in $3$d, the $4$-simplex equation in $4$d, and so on. Being a form of a zero-curvature representation, all these equations are supposed to be related to integrable quantum field theories in the corresponding space-time dimensions. However, it is well known, any non-trivial solution of a $\mathscr{D}$-simplex equation produces an infinite sequence of solutions of $\mathscr{D}-1$-simplex equations and so on. In particular, few known solutions of the quantum tetrahedron equation produce instantly the solutions of the Yang-Baxter equation for the whole $\mathcal{U}_q(A_r)$, $\mathcal{U}_q(B_r)$, $\mathcal{U}_q(C_r)$ and $\mathcal{U}_q(D_r)$ seria in their various representations, and rank $r$ is related to the size of the hidden ``third direction''. So far, there is no place of the higher simplex equations in this scheme, what is curious since the world around us is at least four-dimensional.

The arguments above are given from the ``algebraic'' point of view. Remarkably, quite different but also very sceptic arguments may be given from rather distinct point of view of classical integrable systems and algebraic geometry.

Some solutions to the quantum $4$-simplex equations are known (see \cite{KS} and the next section), but they are trivial in some sense: they do not produce a genuine field theory in four-dimensional space-time, instead they produce a set of field theories in disjoint three-dimensional layers. Modern attempts to find solutions to the $4$-simplex equations seem to produce the similar results \cite{Sotiris}.

Nevertheless, we continue our endeavours to construct non-trivial solutions to the $4$-simplex equations even without understanding of its underlying structures. Recently, in \cite{S} we introduced some idea of how to extend the framework of the pentagon and $10$-term equations \cite{KS} to another framework, it may be called the ``pentagonal algebra'', which potentially may allow one to construct new solutions of the $4$-simplex equations. This paper reports a partial progress in this approach.
We present a few examples of ``structure constants'' for the ``pentagonal algebra'' satisfying a set of specific equations providing solution to the $4$-simplex equation. We do not study the corresponding four-dimensional field theories, this could be the subject of a lengthy separate study.

This paper is organised as follows. In Section 2 we remind the Reader the basic relations between the pentagon equation, $10$-term equation and the $4$-simplex equations. Then, in Sections 3 and 4, we formulate the ``pentagonal algebra equations'', their relation to the $4$-simplex equation, and derive equations for our primary object called the ``Five-leg''. Then, in Sections 5, 6 and 7 we give three examples for these ``Five-legs'' solving all required equations and therefore giving the solutions to the $4$-simplex equations.

\section{Preliminaries: pentagon, 10-terms and 4-simplex}

Many years ago Kashaev and Sergeev \cite{KS} have observed the following. If one defines 
\begin{equation}\label{R4}
R_{0123}^{}\;=\;S_{13}^{} P_{01}^{}P_{23}^{} \overline{S}_{13}^{}\;,
\end{equation}
where $P_{ij}$ is the permutation operator, and substitute it into the $4$-simplex equation
\begin{equation}\label{4s}
R_{0123}R_{0456}R_{1478}R_{2579}R_{3689}\;=\;
R_{3689}R_{2579}R_{1478}R_{0456}R_{0123}\;,
\end{equation}
then a factorisation happens. The $4$-simplex equation decomposes into three disjoint parts: two pentagon equations,
\begin{equation}\label{pent}
S_{23}S_{13}S_{12}\;=\;S_{12}S_{23}\;,\quad
\overline{S}_{12}\overline{S}_{13}\overline{S}_{23}\;=\;\overline{S}_{23}\overline{S}_{12}\;,
\end{equation}
and the $10$-term equation,
\begin{equation}\label{10term}
S_{12} \overline{S}_{13} S_{14} \overline{S}_{24} S_{34} \;=\;
\overline{S}_{24} S_{34} \overline{S}_{14} S_{12} \overline{S}_{13}\;.
\end{equation}
It happened, \emph{for any known} $S_{12}^{}$, satisfying the pentagon equation (left of (\ref{pent})), and for $\overline{S}_{12}^{}\equiv S_{12}^{-1}$, the ten-term relation (\ref{10term}) is satisfied for free.

As it was mentioned in the Introduction, although (\ref{R4}) solves the $4$-simplex equation (\ref{4s}), it does not define a decent quantum field theory in the four-dimensional space-time since the four-dimensional lattice factorises into disjoint three-dimensional sub-lattices described by the $R$-matrix of the tetrahedron equation,
\begin{equation}\label{R3}
R_{123}\;=\;S_{13} P_{23} \overline{S}_{13}\;.
\end{equation}
The integrability structures for this $R$-matrix were studied in \cite{S}, where a generalisation of the form of (\ref{R4}) was suggested.

\section{The generalisation: a ``Five-leg''}

The generalisation of (\ref{R4}) is straightforward:
\begin{equation}\label{R4-2}
R_{0123}\;=\;\sum_a (S_{a})_{13}^{} P_{01}^{} P_{23}^{} (\overline{S}^{\,a})_{13}\;.
\end{equation}
An advantage of this expression is evident: the three-dimensional layers could cling to each other by means of the index `$a$', and thus a hypothetical four-dimensional field theory becomes less trivial.

The next question is: what is the index `$a$' and what are equations for $(S_a)_{12}$ and $(\overline{S}^{\,a})_{12}$? The answer is the following. Equation for $(S_a)_{12}$ is
\begin{equation}\label{PEg}
(S_c)_{23}^{} (S_b)_{13}^{} (S_a)_{12}^{} \;=\; \sum_{d,e} \M_{a,b,c}^{\;d,e} (S_d)_{12}^{} (S_e)_{23}^{}\;,
\end{equation}
while equation for $(\overline{S}^{\,a})_{12}$ is
\begin{equation}\label{nPEg}
(\overline{S}^{\,a})_{12}^{} (\overline{S}^{\,b})_{13}^{} (\overline{S}^{\,c})_{23}^{} \;=\; \sum_{d,e} \overline{\M}^{\,a,b,c}_{\;\,d,e} (\overline{S}^{\,e})_{23}^{} (\overline{S}^{\,d})_{12}^{}\;.
\end{equation}
Finally, the following equation replaces the ten-term one, (\ref{10term}):
\begin{equation}\label{10g}
\begin{array}{l}
\ds \sum_{b_1,b_2,b_3} \overline{\M}^{\, b_1,b_2,b_3}_{\;\,c_1,c_2}
(S_{b_1})_{12}^{} (\overline{S}^{\, a_1})_{13}^{} (S_{b_2})_{14}^{} (\overline{S}^{\,a_2})_{24}^{} (S_{b_3})_{34}^{}\;=\\
\\
\ds \qquad \qquad =\; \sum_{b_1,b_2,b_3}
\M_{b_1,b_2,b_3}^{\;a_1,a_2}
(\overline{S}^{\,b_3})_{24}^{} (S_{c_2})_{34}^{} (\overline{S}^{\,b_2})_{14}^{} (S_{c_1})_{12}^{} (\overline{S}^{\,b_1})_{13}^{}\;.
\end{array}
\end{equation}
Let us call the ``structure constants'' $\M$ and $\overline{\M}$ appeared in the ``pentagonal algebra'' equations (\ref{PEg}-\ref{10g}) the ``Five-legs''.

\section{Equations for the ``Five-legs''}

Equation (\ref{PEg}) implies the self-consistency (or associativity) condition, an analogue of the Jacobi identity for the structure constants $\M_{a,b,c}^{\; d,e}$. It is the following:
\begin{equation}\label{MMM2}
\sum_{b_1,b_2,b_3} 
\M_{a_1,a_2,a_3}^{\;\;\, b_1,b_2}\;
\M_{b_1,a_4,a_5}^{\;\;\, c_1,b_3}\;
\M_{b_2,b_3,a_6}^{\;\;\, c_2,c_3}
\;=\;
\sum_{b_1,b_2,b_3} 
\M_{a_3,a_5,a_6}^{\;\;\, b_2,b_3}\;
\M_{a_2,a_4,b_3}^{\;\;\, b_1,c_3}\;
\M_{a_1,b_1,b_2}^{\;\;\, c_1,c_2}\;.
\end{equation}
The similar equation for $\overline{\M}^{\,a,b,c}_{\;\,d,e}$ has identically the same structure, $\M_{a,b,c}^{\;d,e}\;\to\;\overline{\M}^{\,a,b,c}_{\;\,d,e}$. One has a temptation to consider (\ref{PEg}) as certain algebraic relations, and (\ref{MMM2}) -- literally as the Jacobi identity, and then to think about a ``representation theory''. At least, (\ref{PEg}) has an ``adjoint representation'':
\begin{equation}\label{adj}
\langle i_1,i_2 | S_a | k_1,k_2\rangle \;=\; \M_{a,i_1,i_2}^{\; k_1,k_2}\;,\quad
\langle i_1,i_2 | \overline{S}^{\,a} | k_1,k_2\rangle\;=\; \overline{\M}^{\, a,k_1,k_2}_{\;\, i_1,i_2}\;.
\end{equation}

The additional equation for the ``Five-legs'' is equation (\ref{10g}) in the adjoint representation (\ref{adj}):
\begin{equation}\label{M6}
\begin{array}{l}
\ds \sum_{j_1,\ell_1,j_4,\ell_4 \atop  j_2,j_3,b_1,b_2,b_3}
\overline{\M}_{\;\; c_1,c_2}^{\,b_1,b_2,b_3}\;
\M_{b_1,i_1,i_2}^{\;\,j_1,j_2}\;
\overline{\M}_{\;\; j_1,i_3}^{\,a_1,\ell_1,j_3}\;
\M_{b_2,\ell_1,i_4}^{\;\,k_1,j_4}\;
\overline{\M}_{\;\; j_2,j_4}^{\,a_2,k_2,\ell_4}\;
\M_{b_3,j_3,\ell_4}^{\;\,k_3,k_4}\;=\\
\\
\ds =\; 
\sum_{j_1,\ell_1,j_4,\ell_4 \atop j_2,j_3,b_1,b_2,b_3}
\M_{b_1,b_2,b_3}^{\;\,a_1,a_2}\;
\overline{\M}_{\;\;i_2,i_4}^{\,b_3,j_2,j_4}\;
\M_{c_2,i_3,j_4}^{\;\,j_3,\ell_4}\;
\overline{\M}_{\;\;i_1,\ell_4}^{\,b_2,j_1,k_4}\;
\M_{c_1,j_1,j_2}^{\;\,\ell_1,k_2}\;
\overline{\M}_{\;\;\ell_1,j_3}^{\,b_1,k_1,k_3}\;.
\end{array}
\end{equation}

Note that the definition of the ``adjoint representation'' (\ref{adj}) is not unique.
One can choose different orientations of the ``Five-leg'', and hence equation (\ref{M6}) can be written in different ways\footnote{For instance, I did not study here the possibilities $\langle i_1,i_2 | S_a |k_1,k_2\rangle = \M_{a,k_2,k_1}^{\;i_2,i_1}$, etc.}. For this reason we will call (\ref{MMM2}) the primary equation for the ``Five-leg'' while (\ref{M6}) is an auxiliary equation.

Rewriting (\ref{R4-2}) in the ``adjoint representation'', we come to the
\begin{stat}
If all equations (\ref{MMM2}) and (\ref{M6}) for $\M$ and for $\overline{\M}$ are satisfied, the following expression,
\begin{equation}\label{RMM}
R_{i_0,i_1,i_2,i_3}^{\; k_0,k_1,k_2,k_3}\;=\;
\sum_a \M_{a,i_1,i_3}^{\; k_0,k_2} \, \overline{\M}^{\,a,k_1,k_3}_{\;\, i_0,i_2}\;,
\end{equation}
gives the $R$-matrix for the $4$-simplex\footnote{Moreover, $\ds R_{i_0,i_1,i_2,i_3,i_4}^{k_0,k_1,k_2,k_3,k_4}\;=\;\M_{i_0,i_2,i_4}^{\;\;k_1,k_3}\;\overline{\M}^{\;k_0,k_2,k_4}_{\;\;\;i_1,i_3}$ satisfies the 5-simplex equation.} equation (\ref{4s}).
\end{stat}

It worth to mention in the conclusion of this section, equation (\ref{MMM2}) can also be seen as a set-theoretical equation, and many set-theoretical solutions do exist.

Now we can turn to examples of solutions of (\ref{MMM2},\ref{M6}).

\section{Example 1: a simple ``Five-leg''}

Define an exponent notation,
\begin{equation}\label{exp}
\sk{a;b}\;=\;\left\{
\begin{array}{l}
\ds q^{2ab}\;,\quad a,b\in\mathbb{Z}\;,\\
\\
\ds \omega^{ab}\;,\quad \omega^N=1\;,\;\; a,b\in\mathbb{Z}_N\;,\\
\\
\ds \EXP^{2\pi\ii ab}\;,\quad a,b\in\mathbb{R}\;
\end{array}
\right.
\end{equation}
In general, this is an universal notation implying only $\sk{a;b}=\sk{b;a}$ and $\sk{a;b+c}=\sk{a;b}\sk{a;c}$. 
Also note that 
\begin{equation}\label{gauge}
\M_{a,b,c}^{\; d,e} \;\to\; f_a^{} \,f_b^{} \, f_c^{} \; \M_{a,b,c}^{\; d,e} \; f_d^{-1} \,f_c^{-1}
\end{equation}
is a general gauge transformation of the equation (\ref{MMM2}). It allows one to define Fourier transforms, etc.

\begin{stat}\label{prop1}
Matrix elements of $\M_{a,b,c}^{\;d,e}$, satisfying the equation (\ref{MMM2}), are given by
\begin{equation}\label{M-2}
\M_{a,b,c}^{\;d,e}\;=\;\delta(a+b-d) \delta(b+c-e) \sk{a;c}\;.
\end{equation}
An equivalent formula is the result of the Fourier transform of (\ref{M-2}),
\begin{equation}\label{M-1}
\M_{a,b,c}^{\;d,e}\;=\;\delta(d+e-b) \sk{a-d;c-e}\;.
\end{equation}
Another equivalent formula is the result of the Gauss transform\footnote{Transform with the kernel $G(x,y)=\EXP^{\ii\pi (x-y)^2}$.}  of (\ref{M-2}),
\begin{equation}\label{M-3}
\M_{a,b,c}^{\;d,e}\;=\;\sk{d-e}\sk{a;e-b}\sk{c,d-b}\;.
\end{equation}
For $\sk{a,b}=q^{2ab}$ and cases (\ref{M-1}) and (\ref{M-3}) equation (\ref{MMM2}) converges for $|q|<1$.
\end{stat}
\noindent\textbf{Proof:} direct verification of (\ref{MMM2}). \hfill $\blacksquare$

The origin of the expressions (\ref{M-2}-\ref{M-3}) is the extensive Maple computations for two states which eventually appear to correspond to the middle line in (\ref{exp}). However, I can't state that I found all possible solutions of (\ref{MMM2}) with two states.

\bigskip

Turn now to $R$-matrix (\ref{RMM}). When $\M$ is given by (\ref{M-2}), and when $\overline{\M}^{\,a,b,c}_{\;\,d,e}\;=\;\M_{a,b,c}^{\;d,e}$, equation (\ref{M6}) is satisfied as well. Formula (\ref{RMM}) then produces 
\begin{equation}
R_{i_0,i_1,i_2,i_3}^{\; k_0,k_1,k_2,k_3}\;=\;
\delta(i_1+i_3,k_2)\delta(k_1+k_3,i_2) \delta(i_0+i_1,k_0+k_1) \sk{i_0-k_1,i_3+k_3}\;.
\end{equation}
This is rather simple, but not too trivial $R$-matrix of the $4$-simplex equation. It is non-degenerated\footnote{Inverse $R$ for $\sk{a;b}=\EXP^{2\pi\ii ab}$ is given by $$\overline{R}_{i_0,i_1,i_2,i_3}^{k_0,k_1,k_2,k_3}\;=\;
\delta(i_1+i_3,k_2)\delta(k_1+k_3,i_2) \delta(i_0+i_1,k_0+k_1) \sk{i_1-k_0,i_3+k_3}\;.$$} and not permutation-like. Also, the $4$-simplex equation is non-trivial, it has one Gauss sum/integral in each hand-side (note, for $\sk{a;b}=q^{2ab}$ these sums converge when $|q|<1$, $4$-simplex equation is an equivalence of two $\theta$-constants).

\section{Example 2: another simple ``Five-leg''}

To obtain another elementary ``Five-leg'', one can consider equation (\ref{PEg}) with 
\begin{equation}\label{Saa}
\aa\;=\;(a_0,a_1)\;,\quad S_\aa\;=\;\uop^{a_0}\otimes\vop^{a_1}\;,\quad a_0,a_1\in\mathbb{Z}\;.
\end{equation}
Let for the simplicity the elements $\uop,\vop$ satisfy the Weyl algeba relation,
\begin{equation}\label{Weyl}
\uop\vop\;=\;q^2\vop\uop\;.
\end{equation}
In this case the equation (\ref{PEg}) gives immediately
\begin{equation}\label{M2-1}
\M_{\aa,\bb,\cc}^{\;\dd,\ee}\;=\;
\delta_{a_0+b_0,d_0} \delta_{b_1+c_1,e_1} \delta_{a_1,d_1} \delta_{c_0,e_0} q^{2d_1e_0}\;.
\end{equation}
Equation (\ref{MMM2}) is satisfied in an elementary way since the delta-symbols remove all internal summations.
Next, I suggest an analogous expression for $\overline{\M}$,
\begin{equation}\label{M2-2}
\overline{\M}^{\,\aa,\bb,\cc}_{\;\,\dd,\ee}\;=\;
\delta_{a_0+b_0,d_0} \delta_{b_1+c_1,e_1} \delta_{a_1,d_1} \delta_{c_0,e_0} \;\tilde{q}^{2d_1e_0}\;
\frac{w_{a_0}w_{b_0}}{w_{d_0}} \, \frac{w_{b_1}w_{c_1}}{w_{e_1}}\;.
\end{equation}
This expression differs from (\ref{M2-1}) by a scalar gauge (\ref{gauge}) with $f_{\aa}=w_{a_0}w_{a_1}$, therefore $\overline{\M}$ solves its own equation (\ref{MMM2}).

Turn now to equation (\ref{M6}) for Ansatz (\ref{M2-1},\ref{M2-2}). Two summations remain in each hand side of (\ref{M6}). The left and the right hand sides of (\ref{M6}) coincide if the following identity for $w_a$ is satisfied\footnote{To be precise, there are two identities, one for the «zeroth» component, and another one for the «first» component of $\aa=(a_0,a_1)$. I identify them for simplicity.}:
\begin{equation}\label{wid}
\sum_n \frac{w_{n-a}w_{n-b}w_{c-n}}{w_{n}}\;=\;
\frac{w_{c-a}w_{c-b}}{w_{a}w_{b}}
\sum_n \frac{w_{n-c+a}w_{n-c+b}w_{c-n}}{w_n}\;.
\end{equation}
There is the remarkable observation: this identity is nothing but the so-called ``star-star'' relation for the Ising-type solvable models of lattice statistical mechanics, see e.g. \cite{Baxter}. Since we do not use spectral parameters so far, this equation is satisfied for instance for any Gauss exponent, $w_a=p^{a^2}$ for $a\in\mathbb{Z}$, or $w_a=\EXP^{\ii\pi x^2}$ or $w_a=\EXP^{-a^2}$ for $a\in\mathbb{R}$. Parameters $q$ and $\tilde{q}$, entering (\ref{M2-1}) and (\ref{M2-2}), are not related to each other, they do not enter to summations, they just cancel out.
\\

Thus, one has expressions for $\M$ and $\overline{\M}$, equations (\ref{M2-1},\ref{M2-2}), condition for  $w_a$, (\ref{wid}), all required equations are satisfied, therefore corresponding $R$-matrix (\ref{RMM}) satisfies the $4$-simplex equation (\ref{4s}). \hfill $\blacksquare$

\section{Example 3: less simple ``Five-leg''}

\subsection{Part I}

To start construction of a less simple example, consider again equations (\ref{Saa}) and (\ref{Weyl}). There is another algebra as popular as the Weyl one, this is the $q$-oscillator one:
\begin{equation}\label{qosc}
\uop\vop\;=\;q^2\vop\uop + (1-q^2)\;.
\end{equation}
One has
\begin{stat}\label{prop2}
Let $a,c\in\mathbb{Z}_{\geq 0}$, and $\uop,\vop$ satisfy (\ref{qosc}). Then
\begin{equation}\label{qalg}
\uop^c \vop^a \;=\; \sum_{k=0}^{\min(a,c)} m(a,c;k) \vop^{a-k}\uop^{c-k}\;,\quad
\vop^a \uop^c \;=\; \sum_{k=0}^{\min(a,c)} \overline{m}(a,c;k) \uop^{c-k}\vop^{a-k}\;,
\end{equation}
where
\begin{equation}\label{qm}
m(a,c;k)\;=\;q^{2(a-k)(c-k)} \frac{(q^2;q^2)_{a,c}}{(q^2;q^2)_{k,a-k,c-k}}\;,
\end{equation}
and
\begin{equation}\label{qm2}
\overline{m}(a,c;k)\;=\;(-)^k q^{k(k-1)} q^{-2ac} \frac{(q^2;q^2)_{a,c}}{(q^2;q^2)_{k,a-k,c-k}}\;.
\end{equation}
\end{stat}
\noindent\textbf{Proof:} e.g. the mathematical induction. \hfill $\blacksquare$
 \\
 
\noindent
Two remarkable identities could be mentioned as well:
\begin{stat}
The associativity\footnote{Associativity in a sense $\uop^c (\uop^{c'}\vop^a)=\uop^{c+c'}\vop^a$.} condition for (\ref{qalg}) is
\begin{equation}\label{sum1}
\sum_{k=\max(0,a)}^{\min(b,c)} \frac{q^{2k(k-a)}}{(q^2;q^2)_{k,k-a,b-k,c-k}}\;=\;
\frac{(q^2;q^2)_{b+c-a}}{(q^2;q^2)_{b,c,b-a,c-a}}\;.
\end{equation}
In addition to (\ref{sum1}) we will use one more identity which also holds:
\begin{equation}\label{sum2}
\sum_{k=0}^c x^k \frac{(y;q^2)_{k}(x;q^2)_{c-k}}{(q^2;q^2)_{k,c-k}}\;=\;
\frac{(xy;q^2)_c}{(q^2;q^2)_c}\;.
\end{equation}
\end{stat}
\noindent\textbf{Proof} could be found somewhere in the Gasper-Rahman book \cite{GR}.\hfill $\blacksquare$

Next, we need introduce a function which is related to the Ising-type weight $V_x(n)$ from the paper \cite{BS}. We define \footnote{Bazhanov and Sergeev in \cite{BS} had $\ds V_x(n)=\left(\frac{q}{x}\right)^n\frac{(x^2;q^2)_n}{(q^2;x^2)_n}$, its relation to (\ref{wis}) is evident.}
\begin{equation}\label{wis}
w_\lambda(n)\;=\;\frac{(q^{-2\lambda};q^2)_{n}}{(q^2;q^2)_n}\;,\quad \lambda\in\mathbb{C}\;.
\end{equation}
This function satisfies the ``Star-Star'' relation \cite{BS},
\begin{equation}\label{wid2}
\begin{array}{l}
\ds \frac{w_{\lambda+\nu}(b)}{w_{\lambda+\nu}(b')}\; \sum_n \sk{\lambda;n}\, \frac{w_{\lambda}(n)w_{\mu}(a-n) w_{\nu}(b'-n)}{w_{\lambda+\mu+\nu}(a+b-n)}\;=\;\\
\\
\ds =\;
\sk{\lambda;a-a'}\,
\frac{w_{\lambda+\mu}(a)}{w_{\lambda+\mu}(a')}\;
\sum_n\; \sk{\lambda;n} \frac{w_{\lambda}(n)w_{\mu}(a'-n)w_{\nu}(b-n)}{w_{\lambda+\mu+\nu}(a+b-n)}\;,
\end{array}
\end{equation}
where $a+b=a'+b'$, and we use notation $\sk{a;b}=q^{2ab}$, see (\ref{exp}). 

Next, we have to introduce the notion of the spectral parameters for the "Five-legs''. The spectral parameters are associated with the legs,
\begin{equation}
\aa\;=\;(a_0,a_1)\;\;\to\;\;\biggl( (a_0,\lambda_{a_0}), (a_1,\lambda_{a_1})\biggr)\;.
\end{equation}
Thus there are five pairs of the spectral parameters associated to the ``Five-leg''. They must obey several relations. These relations for $\overline{\M}$ are:
\begin{equation}\label{svz1}
\overline{\M}^{\,\aa,\bb,\cc}_{\;\,\dd,\ee}\;\;:\quad \lambda_{a_0}+\lambda_{b_0}=\lambda_{d_0}\;,\;\;
\lambda_{b_1}+\lambda_{c_1}=\lambda_{e_1}\;,\;\;
\lambda_{a_1}=\lambda_{d_1}\;,\;\;
\lambda_{c_0}=\lambda_{e_0}\;.
\end{equation}
Corresponding relations for $\M_{\aa,\bb,\cc}^{\;\dd,\ee}$ are the same:
\begin{equation}\label{svz2}
\M_{\aa,\bb,\cc}^{\;\dd,\ee}\;\;:\quad \lambda_{a_0}+\lambda_{b_0}=\lambda_{d_0}\;,\;\;
\lambda_{b_1}+\lambda_{c_1}=\lambda_{e_1}\;,\;\;
\lambda_{a_1}=\lambda_{d_1}\;,\;\;
\lambda_{c_0}=\lambda_{e_0}\;.
\end{equation}
Finally, in order to shorten all subsequent notations, we define
\begin{equation}\label{mtr1}
\ds \Phi_{a,b}^{\;c}\;=\;\delta_{a+b,c}\;,\quad \Lambda_{\;a\; ,\; c}^{\;a',\; c'}\;=\;\delta_{a-a',c-c'}\, \frac{\sk{a';c'}}{(q^2;q^2)_{a-a'}}\,
\frac{(q^2;q^2)_{a,c}}{(q^2;q^2)_{a',c'}}\;,
\end{equation}
where $\Lambda$ is just a ``matrix'' notation for (\ref{qm}), and
\begin{equation}\label{mtr2}
\overline{\Phi}^{\,a,b}_{\;\,c}\;=\;
\delta_{c,a+b} \, \frac{w_\lambda(a) w_\mu(b)}{\sk{\lambda;b} w_{\lambda+\mu}(c)}\;,\quad
\overline{\Psi}^{\,a,b}_{\;\,c}\;=\;
\delta_{c,a+b} \, \frac{w_\lambda(a) w_\mu(b)}{\sk{\mu;a} w_{\lambda+\mu}(c)}\;,
\end{equation}
where 
\begin{equation}
\lambda\;=\;\lambda_a\;,\quad \mu\;=\;\lambda_b\;,\quad \lambda+\mu\;=\;\lambda_c
\end{equation}
according to our convention about spectral parameters.
Now we are ready for 
\begin{stat}
Let the ``Five-legs'' are given by 
\begin{equation}\label{Mosc}
\M_{\aa,\bb,\cc}^{\;\dd,\ee}\;=\;
\Phi_{a_0,b_0}^{\;\;d_0}\, \Phi_{b_1,c_1}^{\;\;e_1}\, \Lambda_{a_1,c_0}^{d_1,e_0}\;,
\end{equation}
and
\begin{equation}
\overline{\M}^{\,\aa,\bb,\cc}_{\;\,\dd,\ee}\;=\;
\overline{\Psi}^{\,a_0,b_0}_{\;\; d_0}\; 
\overline{\Phi}^{\,b_1,c_1}_{\;\; e_1}\;
\delta_{a_1,d_1}^{}\, \delta_{c_0,e_0}^{}
\;.
\end{equation}
Then both equations (\ref{MMM2}), for $\M$ and for $\overline{\M}$, are satisfied, and equation (\ref{M6}) is satisfied as well, and therefore the $R$-matrix (\ref{RMM}) satisfies the $4$-simplex equation (\ref{4s}).
\end{stat}
\noindent\textbf{Proof} is the direct verification. However, it requires comments on how all our requirements meet together to provide the final statement. Equation (\ref{MMM2}) for the ``Five-leg'' $\overline{\M}$ is trivial, it coincides with that for the Example 2, see (\ref{M2-2}). Definition of $\M$, eq. (\ref{Mosc}), follows from the definitions (\ref{PEg},\ref{Saa},\ref{qalg}\ref{qm}). Equation (\ref{MMM2}) for this $\M$ factorises into four disjoint factors. Two factors are trivial, they give some delta–functions, while two other factors are two independent copies of the associativity conditions (\ref{sum1}). It has been expected since the equation (\ref{MMM2}) is the associativity condition for (\ref{PEg}). The associativity condition in the matrix form (\ref{mtr1}) is
\begin{equation}\label{as1}
\sum_{a',b',c'}\; \Lambda_{a,\;c}^{a',c'}\; \Lambda_{b,\;c'}^{b',c''}\; \Phi_{a',b'}^{\;\;d}\;=\;
\sum_{d'} \Phi_{a,b}^{\;d'}\; \Lambda_{d',c}^{d,c''}\;.
\end{equation}

Turn now to equation (\ref{M6}). If factorises into six blocks. Two of them are trivial. Two others are equivalent to the identity (\ref{wid2}) with few extra conditions for the spectral parameters. In matrix notations (\ref{mtr2}) they are
\begin{equation}\label{as2}
\sum_e \overline{\Psi}^{e_1,e_2}_{\;\;a}\;
\Phi_{e_1,b}^{\;e_3}\;
\overline{\Psi}^{c,e_4}_{\;\,e_3}\;
\Phi_{e_2,e_4}^{\;\;d}\;=\;
\sum_f \overline{\Psi}^{f_1,f_2}_{\;\;b} \;
\Phi_{a,f_2}^{\;f_3}\;
\overline{\Psi}^{f_4,d}_{\;f_3}\;
\Phi_{f_4,f_1}^{\;\;c}\;,
\end{equation}
and similar for $\overline{\Phi}$. The extra condition for the spectral parameters here are $\lambda_{a}=\lambda_c$, $\lambda_b=\lambda_d$, and $\lambda_{e_2}=\lambda_{f_1}$. These conditions in application to (\ref{M6}) are 
\begin{equation}\label{usl1}
\lambda_{c_{1,0}}\;=\;\lambda_{a_{1,0}}\;,\quad
\lambda_{i_{3,1}}\;=\;\lambda_{k_{2,1}}\;,
\end{equation}
and
\begin{equation}\label{usl2}
\lambda_{j_{4,1}^{}}\;=\;\lambda_{\ell_{4,1}'}\;,\quad
\lambda_{j_{1,0}^{}}\;=\;\lambda_{\ell_{1,0}'}\;,
\end{equation}
where $j_{4,1},j_{1,0}$ are summation indices in the left hand side of (\ref{M6}), and $\ell_{4,1}',\ell_{1,0}'$ are the summation indices in the right hand sides of (\ref{M6}). All the other spectral parameters in all identities (\ref{wid2}) are self-consistent due to the requirements (\ref{svz1},\ref{svz2}).

Finally we turn to two final blocks in the equation (\ref{M6}). In the matrix form,
\begin{equation}\label{as3}
\begin{array}{l}
\ds \sum_{a,b,c'}\;  \overline{\Phi}^{a,b}_{\;d}\; \Lambda_{a,\;c}^{a',c'}\; \Lambda_{b,\;c'}^{b',c''}\;=\;
\sum_{d'}\; \Lambda_{d,\;c}^{d',c''}\; \overline{\Phi}^{a',b'}_{\;\,d'}\;,\\
\\
\ds \sum_{c,b,a'}\; \overline{\Psi}^{c,b}_{\;d}\; \Lambda_{a,\;b}^{a',b'}\; \Lambda_{a',\;c}^{a'',c'}\;=\;
\sum_{d'}\; \Lambda_{a,\;\;d}^{a'',d'}\; \overline{\Psi}^{c',b'}_{\;\,d'}\;.
\end{array}
\end{equation}
Both they are equivalent to the identity (\ref{sum2}). \hfill $\blacksquare$

\subsection{Part II}

The $q$-hypergeometric identities from the previous subsection are known to be just a particular case of identities for a wider class of functions. Such functions are called the ``quantum dilogarithms''. The list of proper quantum dilogarithms can be found in \cite{Kashaev}. 

The most popular example is Faddeev's quantum dilogarithm\footnote{Barnes-Woronovich-Faddev's quantum dilogrithm.} \cite{Fad,FKV} defined by
\begin{equation}
\ph(x)\;=\;\exp\left( \frac{1}{4} \int_{R+\ii 0} \frac{\EXP^{-2\ii x z}}{\sinh (\bl x) \sinh(\bl^{-1}z)} \frac{dz}{z} \right)
\end{equation}
The quantum dilogarithm has the symmetry
\begin{equation}
\ph(x) \ph(-x) \;=\; \phi_0^2 \sk{x}\;,
\end{equation}
where $\phi_0=\ph(0)$ is a constant, and $\sk{x}$ is the Gauss exponent, in the case of Faddeev's dilogarithms the exponents are defined by (see (\ref{exp}))
\begin{equation}
\sk{x}\;=\;\EXP^{\ii\pi x^2}\;,\quad \sk{x;y}\;=\;\EXP^{2\pi\ii xy}\;.
\end{equation}
The crossing parameter $\eta$ in general is defined by
\begin{equation}
\int dy \sk{x;y} \ph(y)\;=\;\gamma\phi_0^2 \frac{\ph(x+\eta)}{\sk{x}}\;,
\end{equation}
where $\gamma$ is another inessential constant\footnote{For Faddeev's dilogarithm $$\phi_0^2=\EXP^{-\ii\pi\eta^2/3-\ii\pi/6}\;,\quad \gamma=\EXP^{\ii\pi/4}\;,\quad \gamma^2\phi_0^6\sk{\eta}\equiv 1\;.$$}. For Faddeev's dilogarithm
\begin{equation}
\eta\;=\;\ii \frac{\bl+\bl^{-1}}{2}\;.
\end{equation}
There is a plenty of identities for the quantum dilogarithms, however we need only one of them:
\begin{equation}\label{sum3}
\begin{array}{l}
\ds \int_{\mathbb{R}+\ii 0} dx \sk{x;-2\eta} \frac{\ph(x+a_1)\ph(x+a_2)}{\ph(x+b_1-\eta)\ph(x+b_2-\eta)}\;=\;\\
\\
\ds =\; \frac{\gamma}{\sk{\eta}} 
\frac{\sk{\eta;b_1+b_2}}{\sk{b_1-b_2}}
\frac{\ph(a_1-b_1)\ph(a_1-b_2)\ph(a_2-b_1)\ph(a_2-b_2)}{\ph(a_1+a_2-b_1-b_2-\eta)}\;.
\end{array}
\end{equation}
This summation formula stays for (\ref{sum1}) and (\ref{sum2}) at the same time.

The final things we need to define before we proceed, is a kernel of operator $\Lambda$ corresponding to (\ref{mtr1}),
\begin{equation}\label{mtr3}
\Lambda_{a,\;c}^{a',c'}\;=\;\delta(a-a'-c+c')\;\frac{1}{\gamma\phi_0^2} \;
\frac{\sk{a';c'}}{\ph(a-a'-\eta)}\;
\frac{\ph(a-\eta)\ph(b-\eta)}{\ph(a'-\eta)\ph(b'-\eta)}\;,
\end{equation}
and a function related to the Boltzmann weight for the Faddeev-Volkov\footnote{The weight for the Fadeev-Volkov model is $\ds V_\lambda(a)\;=\;\sk{\lambda-\eta}\frac{\ph(a-\lambda+\eta)}{\ph(a+\lambda-\eta)}$.} model
\cite{BS-FV}:
\begin{equation}\label{wfv}
w_\lambda(a)\;=\;\frac{\ph(a-\lambda+\eta)}{\ph(a-\eta)}\;.
\end{equation}
It satisfies the ``Star-Star'' relation (\ref{wid2}) with the summation replaced by the integral. The symbols $\overline{\Psi}$ and $\overline{\Phi}$ are given by (\ref{mtr2}) with newly defined $w_\lambda(a)$ and delta-symbols replaced by delta-functions.
Now we are ready to formulate
\begin{stat}\label{prop3}
Let
\begin{equation}\label{vokak}
\begin{array}{l}
\ds \M_{\aa,\bb,\cc}^{\;\dd,\ee}\;=\;
\Phi_{a_0,b_0}^{\;\,d_0}\;
\Phi_{b_1,c_1}^{\;\,e_1}\;
\Lambda_{a_1,c_0}^{d_1,e_0}\;,
\end{array}
\end{equation}
and
\begin{equation}
\begin{array}{l}
\ds \overline{\M}^{\,\aa,\bb,\cc}_{\;\,\dd,\ee}\;=\;
\overline{\Psi}^{a_0,b_0}_{\;\,d_0}\;
\overline{\Phi}^{b_1,c_1}_{\;\,e_1}\;
\delta(a_1-d_1)\; \delta(c_0-e_0)\;.
\end{array}
\end{equation}
The spectral parameters conditions (\ref{svz1},\ref{svz2}) are implied. The additional conditions (\ref{usl1},\ref{usl2}) are implied as well. Then these $\M$ and $\overline{\M}$ satisfiy the equations (\ref{MMM2}) and (\ref{M6}), and, therefore, they produce $R$-matrix (\ref{RMM}) for the $4$-simplex equation \ref{4s}).
\end{stat}
\noindent\textbf{Proof} is a clone of the previous one. In particular, identities (\ref{as1},\ref{as2},\ref{as3}) hold.
 \hfill $\blacksquare$
\\

\noindent
This result requires few comments.
\\

\textbf{Remark 1.} Looking at (\ref{qalg}) and at (\ref{vokak}), one can write the following formula:
\begin{equation}\label{qalg2}
\begin{array}{l}
\ds \sk{\boldsymbol{\alpha};a}\sk{\boldsymbol{\beta};c}=\int \;dx\; m(a,c;x) \;\sk{\boldsymbol{\beta};c-x}
\sk{\boldsymbol{\alpha};a-x}\;,\quad \textrm{where}\\
\\
\ds m(a,c;x)\;=\;\frac{1}{\gamma\phi_0^2} \frac{\sk{a-x;c-x}\ph(a-\eta)\ph(c-\eta)}{\ph(x-\eta)\ph(a-x-\eta)\ph(c-x-\eta)}\;.
\end{array}
\end{equation}
This is an associative multiplication for an algebra generated by operators $\boldsymbol{\alpha},\boldsymbol{\beta}$, generalising in some sense the $q$-oscillator algebra (\ref{qalg}). It seems, such a generalisation of the standard $q$-oscillator algebra was not previously known. Here is another particular example of such algebra, which is is related to  (\ref{qalg2}), but in a more conventional form:
\begin{equation}\label{qalg3}
\begin{array}{l}
\ds 
\uop^a \cdot \vop^c\;=\;
\sum_{n=0}^\infty q^{2(ac+n)} \frac{(q^{2a},q^{2c};q^2)_n}{(q^2;q^2)_n} \;\vop^{c+n}\cdot\uop^{a+n}\;,\\
\\
\ds 
\vop^a\cdot \uop^c \;=\; 
\sum_{m=0}^\infty (-)^n q^{m(m+1)-2(a+m)(c+m)} \frac{(q^{2a},q^{2c};q^2)_m}{(q^2;q^2)_m}\;
\uop^{c+m}\cdot\vop^{a+m}\;.
\end{array}
\end{equation}
The associativity condition for (\ref{qalg3}) is
\begin{equation}
\sum_{n=0}^k q^{2a'n} \frac{(q^{2a};q^2)_n (q^{2a'};q^2)_{k-n}}{(q^2;q^2)_{n,k-n}}\;=\;
\frac{(q^{2(a+a')};q^2)_k}{(q^2;q^2)_k}\;.
\end{equation}
\\

\textbf{Remark 2.} There exists another expression for $\M$,
\begin{equation}\label{vokak2}
\begin{array}{l}
\ds \tilde{\M}_{\aa,\bb,\cc}^{\;\dd,\ee}\;=\;
\delta(a_0+b_0-d_0) \delta(b_1+c_1-e_1)  \delta(a_1-d_1-c_0+e_0)\times\\
\\
\ds \times  \frac{\ph(a_1-d_1+\eta)}{\sk{d_1;e_0}} \;
\frac{\ph(d_1+\eta)\ph(e_0+\eta)}{\ph(a_1+\eta)\ph(c_0+\eta)}\;,
\end{array}
\end{equation}
also satisfying equation (\ref{MMM2}). It is not a result of a gauge transformation of (\ref{vokak}), 
this as an analogue of (\ref{qm2}) for (\ref{qalg2}). For this $\M$ one can also construct corresponding $\overline{\M}$.
\\

\textbf{Remark 3.} Since the gauge group for the dilogarithmic ``Five-leg'' is huge, there are a lot of expressions equivalent to (\ref{vokak}). An example is 
\begin{equation}
\begin{array}{l}
\ds \M_{\aa,\bb,\cc}^{\;\dd,\ee}\;=\;
\frac{\sk{a_1-d_1;e_0-c_0}}{\sk{a_0-d_0;b_0-d_0}\sk{c_1-e_1;b_1-e_1}}\;
\ph(a_1+c_0+\eta)
\times\\
\\
\ds \times  \ph(a_0-d_0+\eta)\ph(b_0-d_0+\eta)\ph(c_1-e_1+\eta)\ph(b_1-e_1+\eta)
\end{array}
\end{equation}
where there are no delta-functions at al.
\\

\textbf{Remark 4.} One may mention an example of a ``trivial'' solution of the Tetrahedron equation, e.g.
\begin{equation}\label{R33}
R_{i_1,i_2,i_3}^{k_1,k_2,k_3}\;=\;\Phi_{i_1,i_3}^{\;\;k_2}\; \overline{\Phi}^{k_1,k_3}_{\;\;i_2}\;,
\end{equation}
where $\Phi$ and $\overline{\Phi}$ are defined by (\ref{mtr1}) and (\ref{mtr2}).
It satisfies the Tetrahedron equation with properly defined spectral parameters, however the corresponding cubic lattice decomposes into a disjoint set of two-dimensional honeycomb lattices. Also, matrix (\ref{R33}) is a subject of the Tetrahedral algebra,
\begin{equation}
R_{123} \; \Lambda_{145} \; \Lambda_{246} \; \Lambda_{356} \;=\; 
\Lambda_{356} \; \Lambda_{246} \; \Lambda_{145} \; R_{123}\;,
\end{equation}
where 
\begin{equation}
\Lambda_{123}\;=\;
\Lambda_{13}\;P_{23}\;,
\end{equation}
with $P_{23}$ -- permutation operator, and $\Lambda_{13}$ is given by either (\ref{mtr1}) or (\ref{mtr3}).

\section{Summary and concluding remarks}

The summary of the results of this paper is the following. Three examples of $R$-matrices for the $4$-simplex equation are presented. Two examples are rather simple while the third one is less simple and moreover, it has a wide degree of generality. A cyclic finite-dimensional case could also be included, in the simplest $N=2$ case the $R$-matrix is just a $256\times 256$ matrix, it deserves a separate study.

The main question left outside this paper is: what a four-dimensional quantum field theory stays behind our solution? 

Also, in the examples here the structure constants $\overline{\M}$ are suspiciously simple. It seems, the possible configurations for the ``adjoint representations'' (\ref{adj}) deserve more detailed investigation, see e.g. the footnote after eq. (\ref{M6}).
\\

\noindent
\textbf{Acknowledgement.} I would like to thank Vladimir Bazhanov, Rinat Kashaev and Vladimir Mangazeev for valuable discussions. 

\end{document}